\documentclass[letter,twocolumn]{jpsj2} %% two-column layout
\title{Thermodynamic Properties and Elementary Excitations in Quantum Sine-Gordon Spin System KCuGaF$_6$}

\author{Rieko {\sc Morisaki}, Toshio {\sc Ono}, Hidekazu {\sc Tanaka}\thanks{E-mail: tanaka@lee.phys.titech.ac.jp.} and Hiroyuki {\sc Nojiri}$^{1}$}

\inst{Department of Physics, Tokyo Institute of Technology, Oh-okayama, Meguro-ku, Tokyo 152-8551\\
$^{1}$Institute for Material Research, Tohoku University, Katahira, Aoba-ku, Sendai 980-8577}

\abst{Thermodynamic properties and elementary excitations in $S=1/2$ one-dimensional Heisenberg antiferromagnet KCuGaF$_6$ were investigated by magnetic susceptibility, specific heat and ESR measurements. Due to the Dzyaloshinsky-Moriya interaction with alternating ${\mib D}$-vectors and/or the staggered $g$-tensor, the staggered magnetic field is induced when subjected to external magnetic field. Specific heat in magnetic field clearly shows the formation of excitation gap, which is attributed to the staggered magnetic field. The specific heat data was analyzed on the basis of the quantum sine-Gordon (SG) model. We observed many ESR modes including one soliton and three breather excitations characteristic of the quantum SG model.
}

\kword{KCuGaF$_6$, one-dimensional antiferromagnet, staggered field, Dzyaloshinsky-Moriya interaction, magnetic-field induced gap, sine-Gordon model, solitons, breathers}

\begin{document}
\maketitle

%\section{Introduction} 
Studies of antiferromagnetic Heisenberg chain (AFHC) have long history. For $S=1/2$ uniform AFHC, the ground state, thermodynamic properties and magnetic excitations are well understood with the help of exact solutions \cite{Hulthen,dCP,Griffiths,Ishimura} and accurate analytical and numerical calculations \cite{Bonner,Eggert,Klumper,Johnston}. These theoretical results demonstrate the importance of the quantum fluctuation characteristic of low-dimensional systems. In particular, exact result of the magnetic excitations in an external magnetic field $H$ is qualitatively different from the results of the linear spin wave theory. The gapless excitations occur at incommensurate wave vectors $q=\pm2{\pi}m(H)$ and ${\pi}\pm2{\pi}m(H)$ in addition to at $q=0$ and ${\pi}$, where $m(H)$ is the magnetization per site in the unit of $g{\mu}_{\rm B}$ \cite{Ishimura}. 

Recently, the physics of $S=1/2$ AFHC in staggered magnetic field induced by the external magnetic field has been attracting considerable attention \cite{Oshikawa1,Affleck}. The model Hamiltonian of such system is written as
\begin{equation}
\mathcal{H}=\sum_{i} \left[J\mib S_i\cdot \mib S_{i+1}-g{\mu}_{\rm B}HS_i^z-(-1)^ig{\mu}_{\rm B}hS_i^x\right], 
\label{eq:model}
\end{equation}
where $h$ is the staggered field perpendicular to the external field $H$ and is given by $h=c_{\rm s}H$. This magnetic model is actualized in some $S=1/2$ AFHC systems such as Cu(C$_6$H$_5$COO)$_2$$\cdot$3H$_2$O \cite{Dender,Asano,Nojiri}, Yb$_4$As$_3$ \cite{Oshikawa2} and PM$\cdot$Cu(NO$_3$)$_2\cdot$(H$_2$O)$_2$ (PM=pyrimidine) \cite{Feyerherm,Zvyagin}. The staggered field originates from the alternating $g$-tensor and the antisymmetric interaction of the Dzyaloshinsky-Moriya (DM) type with the alternating $\mib D$ vector. In these compounds, the field-induced gap ${\Delta}(H)$ almost proportional to $H^{2/3}$ was commonly observed. This field dependence of the gap cannot be explained by the linear spin wave theory, which derives ${\Delta}(H)\propto H^{1/2}$. Using the field theoretical approach, Oshikawa and Affleck \cite{Oshikawa1,Affleck} argued that the model (\ref{eq:model}) can be mapped onto the quantum sine-Gordon (SG) model with Lagrangian density 
$\mathcal{L}=(1/2)\left({\partial}_{\mu}{\phi}\right) + hC\cos (2{\pi}R{\tilde \phi})$,
where $\phi$ is a canonical Bose field, $\tilde \phi$ is the dual field, $R$ is the compactification radius and $C$ is a coupling constant, and that the gap is expressed as ${\Delta}(h)\simeq 1.85J(g{\mu}_{\rm B}h/J)^{2/3}\ln (J/g{\mu}_{\rm B}h)^{1/6}$. Their result is in agreement with experimental results \cite{Dender,Asano,Nojiri,Oshikawa2,Feyerherm,Zvyagin}. 

In the above-mentioned compounds, the exchange interaction is order of 10 K and the proportional coefficient is rather small, $c_{\rm s}=0.08$ \cite{Nojiri,Zvyagin}. For the comprehensive understanding of the systems described by the model (\ref{eq:model}), new compounds having different interaction constants are necessary. In this paper, we show that the static and dynamic properties in a new $S=1/2$ AFHC system KCuGaF$_6$ having a large exchange interaction $J/k_{\rm B}\simeq100$ K can be well described by the model (\ref{eq:model}) with rather large proportional coefficient, $c_{\rm s}\simeq 0.2$, when subjected to external field.

KCuGaF$_6$ crystallizes in a monoclinic structure (space group $P2_1/c$) \cite{Dahlke}. 
Cu$^{2+}$ and Ga$^{3+}$ ions surrounded octahedrally by six F$^-$ ions form a pyrochlore lattice. Since Cu$^{2+}$ ions with spin 1/2 are arranged almost straightforward along the $c$-axis and neighboring Ga$^{3+}$ ions are nonmagnetic, the exchange interaction between neighboring Cu$^{2+}$ ions should have one-dimensional (1D) nature. CuF$_6$ octahedra are elongated perpendicular to the chain direction parallel to the $c$-axis due to the Jahn-Teller effect. The hole orbitals of Cu$^{2+}$ ions are linked along the chain direction through the $p$ orbitals of F$^-$ ions with bond angle of 129$^\circ$ for Cu$^{2+}-$F$^--$Cu$^{2+}$. This orbital configuration can generate strong antiferromagnetic exchange interaction along the chain. The elongated axes are alternate along the chain direction. The local principal axis of each octahedron is tilted from the $c$-axis by $\pm 25.4^\circ$, which leads to the staggered $g$-tensor. At present, details of the $g$-tensor is not clear, because no ESR signal at X ($\sim 9$ GHz) and K ($\sim 24$ GHz) bands frequencies is observed at room temperature due to large linewidth. For the $\mib D$-vector of the DM interaction, the component parallel to the $ac$-plane alternates along the chain direction, because there is the $c$-glide plane at $\pm b/4$. 

%\section{Experimental}
Single KCuGaF$_6$ crystals were grown by the vertical Bridgman method from the melt of equimolar mixture of KF, CuF$_2$ and GaF$_3$ sealed in Pt-tube. The temperature at the center of the furnace was set at 750 $^\circ$C, and the lowering rate was 3 mm/h. The materials were dehydrated by heating in vacuum at about 150$^{\circ}$C for three days. Transparent colorless crystals with typical size $3 \times 3\times 3$ mm$^3$ were obtained. Crystals obtained were identified to be KCuGaF$_6$ by X-ray powder and single crystal diffractions. KCuGaF$_6$ crystals were cleaved along the $(1, 1, 0)$ plane. Magnetic susceptibilities were measured using a SQUID magnetometer (Quantum Design MPMS XL) down to 1.8 K. Specific heat measurements were carried out down to 0.46K in magnetic fields of up to 9 T using a physical property measurement system (Quantum Design PPMS) by the relaxation method. The high-frequency, high-field ESR
measurement was performed in the frequency range 66.5-761.6 GHz using the terahertz electron spin resonance apparatus (TESRA-IMR) \cite{Nojiri} at the Institute for Material Research, Tohoku University. Magnetic fields up to 30 T was applied by a multilayer pulse magnet. FIR lasers, backward traveling wave tubes and Gunn oscillators were used as light sources.  

\begin{figure}[htbp]
  \begin{center}
    \includegraphics[keepaspectratio=true,width=60mm]{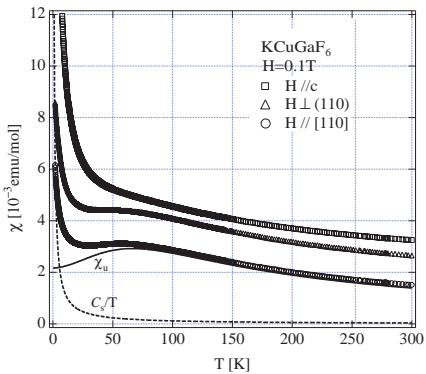}
  \end{center}
  \caption{Temperature dependences of magnetic susceptibilities $\chi$ in KCuGaF$_6$ measured at $H=0.1$ T for $H\parallel c$, $H\perp(1,1,0)$ and $H\parallel [1,1,0]$. The values of the susceptibilities are shifted upward consecutively by $1\times 10^{-3}$ emu/mol. The solid and dashed lines denote $\chi_{\rm u}$ and $C_{\rm s}/T$ for $H\parallel [1,1,0]$, respectively.}
  \label{fig:susceptibility}
\end{figure}

%\section{Results and Discussion}
Figure \ref{fig:susceptibility} shows the temperature dependences of the magnetic susceptibilities $\chi$ measured at $H$=0.1 T for $H\parallel c$, $H\perp (1, 1, 0)$ and $H\parallel [1, 1, 0]$. The differences between the absolute values of the three susceptibilities for $T > 100$ K are due to the anisotropies of the $g$-factor. For $H\perp (1,1,0)$ and $H\parallel [1,1,0]$, the susceptibilities exhibit broad maxima at $T\sim 70$ K characteristic of $S=1/2$ AFHC. With decreasing temperature, the susceptibilities increase rapidly below 20 K, obeying the Curie law. For $H\parallel c$, the Curie term is so large that the broad susceptibility maximum is hidden. The Curie term is intrinsic to the present system, because its magnitude depends on field direction and is independent of sample.

According to the theory by Oshikawa and Affleck \cite{Oshikawa1,Affleck}, the magnetic susceptibility for the model (\ref{eq:model}) is expressed as $\chi \simeq \chi_{\rm u} + c_{\rm s}^2\chi_{\rm s}$, where $\chi_{\rm u}$ is the uniform susceptibility for $S=1/2$ AFHC without anisotropy \cite{Eggert,Johnston} and $\chi_{\rm s}$ is the staggered susceptibility given by 
\begin{equation}
\chi_{\rm s}(T)\simeq 0.278\left(\frac{N_{\mathrm{A}}g^2\mu_{\mathrm{B}}^2}{4k_{\mathrm{B}}T}\right)\left \{ \ln \left(\frac{J}{k_{\mathrm{B}}T}\right)\right \}^{1/2}.
\label{eq:chi_s}
\end{equation}
The staggered susceptibility $\chi_{\rm s}$ has a tendency to diverge for $T\rightarrow 0$. Thus, the Curie term in the magnetic susceptibility of KCuGaF$_6$ can be attributed to the staggered susceptibility $\chi_{\rm s}$ due to the staggered field $h$. The presence of the Curie term in magnetic susceptibility is consistent with the model (\ref{eq:model}). However, the Curie term observed has a long tail to the high-temperature region and is not describable by $\chi_{\rm s}$ of eq. (\ref{eq:chi_s}). 
Since the subleading logarithmic term of $\chi_{\rm s}$ is valid only for $T\ll J/k_{\rm B}$ and there is no analytical result on the subleading term that is applicable for $T\sim J/k_{\rm B}$, we use ${\chi}= \chi_{\rm u}+C_{\rm s}/T$ to analyze experimental susceptibilities. Fitting this equation to the susceptibility data, we obtained exchange constant $J/k_{\mathrm{B}}$ and the Curie constant $C_{\rm s}$. The $g$-factors used are $g=2.18$, 2.32 and 2.33 for $H\parallel c$, $H\perp (1,1,0)$ and $H\parallel [1,1,0]$, respectively, which were determined by the present ESR measurements at $T\sim 60$ K. The exchange constant can be determined accurately when the staggered susceptibility is small. Thus, we evaluate the exchange constant as $J/k_{\rm B}=103\pm 1$ K from the average of $J/k_{\rm B}$ for $H\perp (1, 1, 0)$ and $H\parallel [1, 1, 0]$. The Curie constant $C_{\rm s}$ for the staggered susceptibility is strongly dependent on the field direction and is the largest for $H\parallel c$. This result indicates that the staggered field induced by external field is the largest for $H\parallel c$.
\begin{figure}[htbp]
  \begin{center}
    \includegraphics[keepaspectratio=true,width=60mm]{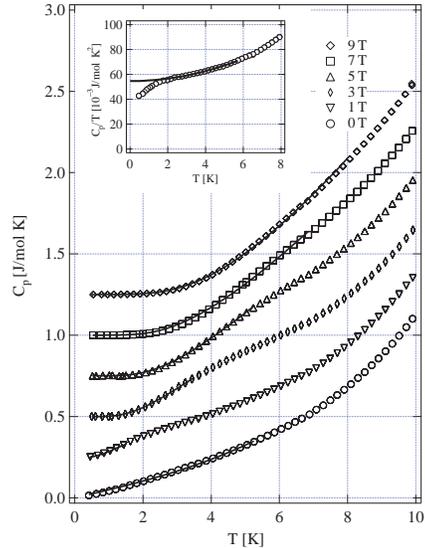}
  \end{center}
  \caption{Total specific heat $C_{\rm p}$ in KCuGaF$_6$ measured at various magnetic fields for $H\parallel c$. The values of the specific heat are shifted upward consecutively by 0.25 J/mol K. The inset shows $C_{\rm p}/T$ vs $T$ measured at zero magnetic field. Solid line for $H=0$ is fit by $C_{\rm p}={\gamma}T+bT^3$ and those for $H\neq 0$ are fits by $C_{\rm p}=C_{\rm SG}+bT^3$. Specific heat data for $T < 5$ K were thinned out so that fitting curves are visible.}
  \label{fig:heat}
\end{figure}

Figures \ref{fig:heat} shows the total specific heat $C_{\rm p}$ measured at various magnetic fields for $H\parallel c$. No magnetic ordering was observed down to 0.46 K, which indicates good one-dimensionality of the present system. We estimate the upper limit of interchain interaction $J'$ in KCuGaF$_6$ as $J'/J < 2\times 10^{-3}$, using eq. (8) in ref. \citen{Yasuda}. 

Specific heat at zero field exhibits linear temperature dependence below 4 K characteristic of the $S=1/2$ HAFC \cite{Bonner,Klumper,Johnston}. Low-temperature specific heat at zero field is described as
$C_{\rm p}(T)={\gamma}T+bT^3$ with ${\gamma}=2Rk_{\rm B}/(3J)$, where the first and the second terms denote the magnetic and lattice contributions, respectively. Fitting this equation to specific heat data at $H=0$ for $T < 5$ K, we obtain $J/k_{\rm B}=103\pm 2$ K and $b=(5.4\pm 0.3)\times 10^{-4}$ J/mol K$^{3}$. This exchange constant coincides with $J/k_{\rm B}=103$ K evaluated from magnetic susceptibility data. The inset of Fig. \ref{fig:heat} shows $C_{\rm p}/T$ vs $T$ measured at zero magnetic field. The proportional coefficient $\gamma$ decreases gradually below 1.5 K. The existence of residual magnetic field trapped in the superconducting magnet is not responsible for this phenomenon, because it does not depend on sample orientation. This phenomenon cannot also be attributed to the logarithmic correction due to the critical quantum fluctuation \cite{Klumper,Johnston}, because this effect should occur at lower temperatures close to $T=0$. At present, the origin leading to the decrease in $\gamma$ below 1.5 K is not clear.

In finite magnetic field, low-temperature specific heat $C_{\rm p}$ exhibits exponential dependence on temperature, which is enhanced with increasing magnetic field. This result clearly indicates that the magnetic field induces the excitation gap, and that the gap increases with magnetic field. In what follows, we analyze the specific heat data using the results of the quantum SG field theory \cite{Oshikawa1,Affleck,Essler1}. 

In the quantum SG model, elementary excitations are solitons and breathers. The breathers are soliton-antisoliton bound states. Essler {\it et al.} \cite{Essler1} calculated the field dependence of the soliton mass $M_{\rm s}$ corresponding to the excitation energy at $q=\pm 2{\pi}m(H)$ and ${\pi}\pm 2{\pi}m(H)$. Their result is expressed as
\begin{equation}
M_{\rm s}=\frac{2Jv}{\sqrt{\pi}}\frac{\Gamma \left(\displaystyle\frac{\xi}{2}\right)}{\Gamma \left(\displaystyle\frac{1+\xi}{2}\right)} \left[\frac{\Gamma \left(\displaystyle\frac{1}{1+\xi}\right)}{\Gamma \left(\displaystyle\frac{\xi}{1+\xi}\right)} \frac{c{\pi}g{\mu}_{\rm B}H}{2Jv}c_{\rm s}\right]^{(1+\xi)/2},
\label{eq:solitonmass}
\end{equation}
where $v$ is the dimensionless spin velocity, $\xi$ is a parameter given by 
${\xi}=[2/({\pi}R^2)-1]^{-1}$
and $c$ is a parameter depending on magnetic field. The field dependences of these parameters are shown in literature \cite{Affleck,Essler1}. Here, $v\rightarrow {\pi}/2$, ${\xi}\rightarrow 1/3$ and $c\rightarrow 1/2$ for $H\rightarrow0$.
The mass of the $n$-th breather, $M_n$, which corresponds to the excitation energy at $q=0$ and ${\pi}$, is given by 
\begin{equation}
M_n=2M_{\rm s} \sin \left(\frac{n{\pi}{\xi}}{2}\right), \hspace{0.5cm}n=1, \cdots , \left[{\xi}^{-1}\right], 
\label{eq:breather}
\end{equation}
where $n$ is integer and its upper limit is determined by ${\xi}^{-1}$ \cite{Affleck}. In our experimental field range, breathers up to the third order can exist. At low temperatures compared with $M_{\rm s}/k_{\rm B}$, specific heat due to solitons and breathers is given by 
\begin{eqnarray}
C_{\rm SG} &\sim& \sum_{\alpha=0}^{[1/ \xi]}\frac{(1+\delta_{\alpha 0})M_{\alpha}R}{\sqrt{2\pi}Jv}\left[ 1+\frac{k_{\mathrm{B}}T}{M_{\alpha}}+\frac{3}{4}\left( \frac{k_{\mathrm{B}}T}{M_{\alpha}}\right)^2\right] \nonumber  \\
& &  \times \left( \frac{M_{\alpha}}{k_{\mathrm{B}}T} \right)^{3/2} \exp\left({-\frac{M_{\alpha}}{k_{\mathrm{B}}T}}\right), 
 \label{eq:C_SG}
\end{eqnarray}
where $M_0\equiv M_{\rm s}$ \cite{Troyer,Essler2,Chen}. Since there exist one soliton, one antisoliton and three breathers for $0\leq q < {\pi}$, the soliton contribution is doubly counted in $C_{\rm SG}$. To obtain the soliton mass $M_{\rm s}$, we fit $C_{\rm p}=C_{\rm SG}+bT^3$ to the specific heat data for $T < 0.4M_{\rm s}/k_{\rm B}$. The soliton mass obtained for $H\parallel c$ is plotted as a function of the external field in Fig. 3. Setting $J/k_{\rm B}=103$ K and $c_{\rm s}=0.17\pm 0.01$ in eq. (\ref{eq:solitonmass}), we can achieve a good fit to the experimental soliton mass, as shown by the solid line in Fig. 3. 
\begin{figure}[tb]
\begin{center}
\includegraphics[keepaspectratio=true,width=60mm]{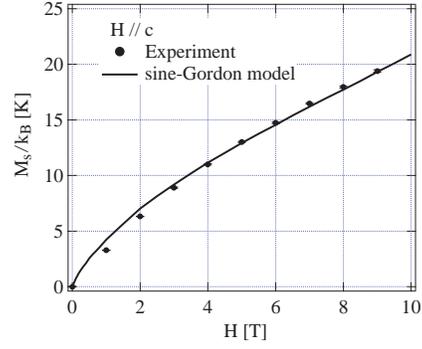}
\caption{Soliton mass $M_{\rm s}$ obtained by specific heat measurements for $H\parallel c$ in the unit of K. Solid line denotes the soliton mass calculated from eq. (\ref{eq:solitonmass}) with $J/k_{\rm B}=103$ K and $c_{\rm s}=0.17$.}
\end{center}
\label{solitonmass}
\end{figure}

\begin{figure}[tb]
\begin{center}
\includegraphics[keepaspectratio=true,width=60mm]{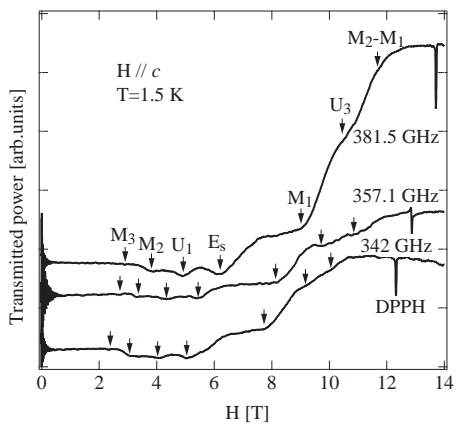}
\end{center}
\caption{Examples of ESR spectra in KCuGaF$_6$ measured at 1.5 K for $H\parallel c$. Arrows indicate resonance fields.}
\label{fig:ESR_spectra}
\end{figure}

\begin{figure}[tb]
\begin{center}
\includegraphics[keepaspectratio=true,width=62mm]{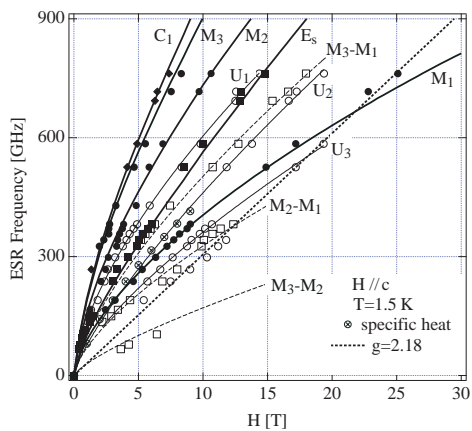}
\end{center}
\caption{The frequency-field diagram of the ESR modes in KCuGaF$_6$ measured at 1.5 K for $H\parallel c$. Symbols denote experimental results and thick solid lines labeled as $E_{\rm s}, M_1, M_2$ and $M_3$ are resonance conditions calculated from eqs. (\ref{eq:solitonmass}), (\ref{eq:breather}) and (\ref{eq:soliton_resonance}) with $J/k_{\rm B}=103$ K and $c_{\rm s}=0.17$. Soliton mass obtained from specific data are also plotted.}
\label{fig:ESR_diagram}
\end{figure}

To observe elementary excitations at $q=0$ in KCuGaF$_6$, we performed high-frequency, high-field ESR measurements. Figure \ref{fig:ESR_spectra} shows examples of ESR spectra obtained at 1.5 K for $H\parallel c$. Absorption signals observed both on sweeping field up and down were determined as intrinsic resonance signals. Many resonance modes were observed. The resonance data are summarized in Fig. \ref{fig:ESR_diagram}. For ESR, the quantum SG field theory predicts one soliton resonance given by
\begin{equation}
E_{\rm s}=\sqrt{M_{\rm s}^2+(g\mu_{\mathrm{B}}H)^2},
\label{eq:soliton_resonance}
\end{equation}
and three breather excitations given by eq. (\ref{eq:breather}) \cite{Affleck}. The soliton resonance corresponds to the $q=0$ excitation on the excitation branch connected to the solitons at $q=\pm 2{\pi}m(H)$. Taking into consideration of the field dependences of the excitation energies for soliton and breathers evaluated from specific heat measurements, we assign resonance modes labeled by $E_{\rm s}, M_1, M_2$ and $M_3$ as the soliton resonance and the first, second and third breathers given by eqs. (\ref{eq:soliton_resonance}) and (\ref{eq:breather}), respectively. The intensities of the soliton resonance $E_{\rm s}$ and the first breather $M_1$ are of the same order. The soliton resonance occurs when the oscillating magnetic field $H_1$ is perpendicular to the external field $H$, while breathers are excited for $H_1\parallel H$. In the present experiments, the light propagates in a light pipe parallel to the external magnetic field with several propagation modes, so that the oscillating magnetic field has both components parallel and perpendicular to the external field. Thus, both soliton resonance and breathers can be observed.
 
Adjustable parameter to describe the resonance condition is the proportional coefficient $c_{\rm s}$ only. Thick solid lines in Fig. \ref{fig:ESR_diagram} are the resonance conditions of the soliton resonance and breathers calculated from eqs. (\ref{eq:solitonmass}), (\ref{eq:breather}) and (\ref{eq:soliton_resonance}) with $J/k_{\rm B}=103$ K and $c_{\rm s}=0.17\pm 0.01$. The experimental results are successfully described by the quantum SG field theory. The proportional coefficient $c_{\rm s}=0.17$ determined by the present ESR measurements coincides with $c_{\rm s}=0.17$ determined by the specific heat measurements. These results indicate that the thermodynamic properties and elementary excitations in KCuGaF$_6$ are consistently described by the quantum SG model.

There are additional modes labeled as $M_2-M_1, M_3-M_1$ and $M_3-M_2$ in Fig. \ref{fig:ESR_diagram}. Their excitation energies correspond to the energy differences between two in three breathers, as shown by thin dashed lines. Hence, these excitations can be assigned as the inter-breather transitions. In the present pulsed high magnetic field, excitation levels split under almost adiabatic condition, keeping the population at zero magnetic field. Thus, such inter-breather transitions are observable. The $C_1$ mode with the lowest resonance field is considered to be the multiple excitations of the soliton resonance and the first breather, because its energy is equal to $E_{\rm s}+M_1$. At present, we do not have clear explanation of $U_1$ to $U_3$ modes denoted by thin solid lines.

%\section{Conclusion}
In conclusion, we have presented the results of magnetic susceptibility, specific heat and ESR measurements on the $S=1/2$ AFHC KCuGaF$_6$ with the large exchange interaction $J/k_{\rm B}=103\pm 2$ K. Due to the DM interaction with alternating ${\mib D}$-vectors and/or the staggered $g$-tensor, the staggered magnetic field is induced when external magnetic field is applied. Thus, the effective magnetic model in magnetic field can be expressed by the quantum SG model. Although small disagreement was observed in magnetic susceptibility, most experimental results were beautifully described by the quantum SG field theory. For $H\parallel c$, the proportional coefficient was obtained as $c_{\rm }=0.17\pm 0.01$. KCuGaF$_6$ differs from other quantum SG systems in large exchange interaction and proportional coefficient. Hence, KCuGaF$_6$ is suitable for understanding elementary excitations at relatively low-field region.

%\section*{Acknowledgment}
This work was supported by a Grant-in-Aid for Scientific Research from the Japan Society for the Promotion of Science and also supported by a 21st Century COE Program at Tokyo Tech ``Nanometer-Scale Quantum Physics'' and a Grant-in-Aid for Scientific Research on Priority Areas ``High Field Spin Science in 100 T'' (No. 451) both from the Ministry of Education, Culture, Sports, Science and Technology.

\end{document}